# Large-kernel convolutional neural networks for wide parameter-space searches of continuous gravitational waves

Prasanna M. Joshi[*] and Reinhard Prix

*Max Planck Institute for Gravitational Physics (Albert-Einstein-Institute), 30167 Hannover, Germany
and Leibniz Universität Hannover, 30167 Hannover, Germany*

The sensitivity of wide parameter-space searches for continuous gravitational waves (CWs) is limited by their high computational cost. Deep learning is being studied as an alternative method to replace various aspects of a CW search. In previous work [Phys. Rev. D **108**, 063021 (2023)], new design principles were presented for deep neural network (DNN) search of CWs and such DNNs were trained to perform a targeted search with matched-filtering sensitivity. In this paper, we adapt these design principles to build a DNN architecture for wide parameter-space searches in 10 days of data from two detectors (H1 and L1). We train a DNN for each of the benchmark cases: six all-sky searches and eight directed searches at different frequencies in the search band of 20–1000 Hz. We compare our results to the DNN sensitivity achieved from Dreissigacker and Prix [Phys. Rev. D **102**, 022005 (2020)] and find that our trained DNNs are more sensitive in all the cases. The absolute improvement in detection probability ranges from 6.5% at 20 Hz to 38% at 1000 Hz in the all-sky cases and from 1.5% at 20 Hz to 59.4% at 500 Hz in the directed cases. An all-sky DNN trained on the entire search band of 20–1000 Hz shows a high sensitivity at all frequencies, providing a proof of concept for training a single DNN to perform the entire search. We also study the generalization of the DNN performance to signals with different signal amplitude, frequency, and the dependence of the DNN sensitivity on sky position.

## I. INTRODUCTION

Continuous gravitational waves (CWs) are very weak, long-lasting, and quasimonochromatic waves that are emitted by rapidly spinning neutron stars with a nonaxisymmetric perturbation. Due to the very small amplitude of these waves, a significant detection requires a search over months/years of data to gain enough signal power. Several searches have been performed on data from LIGO (H1 and L1) and Virgo (V1) detectors, but no CWs have yet been detected [1].

The coherent matched filter is the most sensitive search method for CWs that involves cross correlating the detector data with several templates of signals over the entire search time span coherently. But this method is infeasible for long time spans of months/years as it has a very high computational cost for wide parameter-space searches because of the large number of signal templates, e.g., see [2]. Instead, searches are performed using a semicoherent matched-filter method in which short segments of the total time span are cross correlated coherently and their results are combined incoherently. This method is more sensitive than the coherent method at a fixed computational cost.

An alternative method for reducing the computational cost with minimal loss of sensitivity is to train a deep neural network (DNN) to perform a search. There have been a number of studies exploring the potential of DNNs to help improve CW searches, for example, as a clustering and follow-up method of search candidates [3–5], to reduce the computational cost of follow-ups [6,7], and to mitigate the effect of instrumental noise artifacts [8]. DNNs have also been shown to be able to accelerate searches for long-duration, transient CWs [9–11].

The approach taken in [12–14] and in this paper is to train a DNN directly on detector strain data to identify a CW signal. In our previous work [12], we have presented new design principles for DNNs to perform CW searches. We found that a DNN with an architecture based on these principles can be trained to be as sensitive as a coherent matched filter in a targeted search. In this paper, we train DNNs with architecture motivated by these principles to perform all-sky searches and directed searches for CWs emitted by an isolated neutron star. We compare our search sensitivity with a similar DNN search and a coherent matched-filter search characterized in [13], and demonstrate a substantial overall improvement over the previous DNN sensitivity.

---

[*]Contact author: prasanna.mohan.joshi@aei.mpg.de

This paper is organized as follows: In Sec. II, we introduce the all-sky and directed search benchmarks, and in Sec. III, we describe the architecture of the DNN, the training process, and calculation of test metrics. We present our test results and their comparison with earlier results in Sec. IV and the conclusions and future work in Sec. V.

## II. COMPARISON TEST BENCHMARKS

We characterize DNNs as a search method on the following benchmarks: an all-sky search and two directed searches pointing at Supernova remnants Cassiopeia A (CasA) and G347.3-0.5 (G347). We use the same parameters for these searches as in [13] (listed in Tables I and II), except that we define our searches with a time span of 10 days instead of $10^6$ s (∼11.6 days). This allows us to compare our results with a coherent matched-filter search using WEAVE [15] as well as results of the DNN search in [13] after appropriate rescaling.

We measure the sensitivity of a search method by calculating detection probability $p_{\text{det}}$ at fixed false-alarm probability $p_{\text{fa}} = 1\%$ per 50 mHz frequency band. The $p_{\text{fa}}$ sets a threshold on the value of detection statistic for a search performed over a frequency band of 50 mHz.

TABLE I. Definition of all-sky benchmark searches.

| | |
|---|---|
| Start time | 1 200 300 463 s |
| Duration | 10 days |
| Detectors | LIGO Hanford (H1) and Livingston (L1) |
| Noise | Stationary, white, Gaussian |
| $\tau_{\text{ref}}$ | 1 200 300 463 s |
| Sky region | All sky |
| Frequency band | $f \in [20, 1000]$ Hz |
| Spin-down range | $\dot{f} \in [-10^{-10}, 0]$ Hz s$^{-1}$ |

TABLE II. Definition of directed benchmark searches, modeled after [16].

| | |
|---|---|
| Start time | 1 200 300 463 s |
| Duration | 10 days |
| Detectors | LIGO Hanford (H1) and Livingston (L1) |
| Noise | Stationary, white, Gaussian |
| $\tau_{\text{ref}}$ | 1 200 300 463 s |
| Sky position | CasA/G347 |
| Frequency band | $f \in [20, 1000]$ Hz |
| Spin-down range | $-f/\tau \leq \dot{f} \leq 0$ Hz s$^{-1}$ |
| Second order spin-down | $0 \leq \ddot{f} \leq 5f/\tau^2$ Hz s$^{-2}$ |
| Characteristic age ($\tau$) | CasA: 330 yr, G347: 1600 yr |

The $p_{\text{det}}$ is computed for a population of signals injected at a fixed signal amplitude $h_0$.

The sensitivity can also be characterized as an *upper-limit* amplitude $h_0^{p_{\text{det}}}$ at which a search has a given detection probability $p_{\text{det}}$ at a fixed $p_{\text{fa}}$. This upper-limit amplitude $h_0^{p_{\text{det}}}$ scales with the amplitude spectral density $\sqrt{S_n}$ of the noise at every frequency [17]. Therefore, it is more convenient to define *sensitivity depth* $\mathcal{D}$ as follows:

$$\mathcal{D} \equiv \frac{\sqrt{S_n}}{h_0}, \qquad (1)$$

The sensitivity depth $\mathcal{D}^{90\%}$ corresponds to the upper-limit signal amplitude $h_0^{90\%}$, for which a search would achieve a $p_{\text{det}} = 90\%$ at a fixed $p_{\text{fa}} = 1\%$ per 50 mHz bandwidth.

Similarly, we can calculate $p_{\text{det}}$ for a population of signal injections at a fixed value of signal power $\rho^2$, which is defined as follows (from [17]):

$$\rho^2 \equiv \frac{4}{25} \frac{T_{\text{data}}}{\mathcal{D}^2} R^2(\theta), \qquad (2)$$

where $T_{\text{data}}$ is the total duration of data from all the detectors and $R(\theta)$ is a geometric antenna response factor that depends on signal parameters $\theta = \{\alpha, \delta, \cos\iota, \psi\}$, where $(\alpha, \delta)$ is the source sky position, $\psi$ is the polarization angle, and $\iota$ is the angle between the neutron star rotation axis and the line of sight.

The coherent matched-filter searches in [13] are performed using WEAVE on simulated data by injecting signals into Gaussian noise at a fixed $\mathcal{D}$. In order to keep the computational cost of the search low, it was performed on narrow frequency bands of 50 mHz with different starting frequencies in the entire search band. In our searches, we also select a 50 mHz band with the same starting frequencies as those used in [13] for a fair comparison. The sensitivity of these searches was characterized by calculating the $\mathcal{D}^{90\%}$ for each of the representative cases as shown in Table III.

We cannot directly use these values of $\mathcal{D}^{90\%}$ to compare the sensitivity of the WEAVE search with our DNN search due to their different time spans. As the senstivity of a search depends only on $\rho^2$ of a signal (see [17]), we can apply Eq. (2) at a constant $\rho^2$ and rescale the $\mathcal{D}^{90\%}$ with $T_{\text{data}}$ as $\mathcal{D}^{90\%} \sim \sqrt{T_{\text{data}}}$. These rescaled WEAVE $\mathcal{D}^{90\%}$ values for our search cases are given in Table III. Certain cases in the directed searches do not have a $\mathcal{D}^{90\%}$ value calculated directly from the WEAVE search. We estimate these intermediate values by linearly interpolating between the known values and then rescaling them as described above.

## III. DEEP LEARNING

In this section, we describe the architecture of the DNN, preprocessing of input data, the training process, and the metrics that we use to evaluate the performance of the trained DNNs. The design of the DNN presented here is based on the design principles in Sec. IV A of [12]. We have made appropriate changes to the design from [12] in order to optimize the architecture to better be able to learn a large number of signal shapes in a wide parameter-space search.

TABLE III. Sensitivity depths $\mathcal{D}^{90\%}$ (at $p_{\mathrm{fa}} = 1\%$ per 50 mHz) achieved by the WEAVE coherent matched-filtering search described in [13] for a time span of $10^6$ s and the corresponding $\mathcal{D}^{90\%}$ rescaled to a time span of 10 days for our benchmarks using Eq. (2). The $\mathcal{D}^{90\%}$ for the directed search cases for which corresponding values have not been calculated in [13] (in bold), are computed by first estimating $\mathcal{D}^{90\%}$ for a $10^6$-s search using linear interpolation between known values and then rescaling to 10 days according to Eq. (2). The DNNs are trained on data containing signal injections at the values of $\mathcal{D}$ from this table.

| $\mathcal{D}^{90\%}_{\mathrm{MF}}$ ($/\sqrt{\mathrm{Hz}}$) | All sky | | Directed (G347) | | Directed (CasA) | |
|---|---|---|---|---|---|---|
| | This work ($T_{\mathrm{span}} = 10$ days) | Prev. work ($T_{\mathrm{span}} = 10^6$ s) | This work ($T_{\mathrm{span}} = 10$ days) | Prev. work ($T_{\mathrm{span}} = 10^6$ s) | This work ($T_{\mathrm{span}} = 10$ days) | Prev. work ($T_{\mathrm{span}} = 10^6$ s) |
| 20 Hz | 39.0 | 42.0 | 42.9 | 46.1 | 42.9 | 46.1 |
| 100 Hz | 37.3 | 40.1 | **42.5** | … | **42.5** | … |
| 200 Hz | 36.6 | 39.4 | **42.1** | … | **42.0** | … |
| 500 Hz | 35.6 | 38.3 | 40.8 | 43.9 | … | 43.8 |
| 1000 Hz | 33.4 | 35.9 | **40.4** | … | … | 43.4 |
| 1500 Hz | … | … | … | 43.1 | … | … |

### A. Preprocessing the input

We convert the one-dimensional time-series data for each detector into a two-channel (real and imaginary part) short-time Fourier transform (STFT) by dividing the time series into a number of segments. We divide the time span of 10 days into 40 segments of 6 h each, which is in contrast with 10 segments of one day each used in [12]. Thus, 4 consecutive segments represent the daily repeating diurnal frequency pattern illustrated in Fig. 1 of [12] and supports the application of convolutional layers to learn these repeating patterns. This change, which is consistent with the design principles outlined in [12], was made empirically by testing DNN performance for different values of segment duration ranging from three hours to one day.

The spectrograms from the two detectors (H1 and L1) are stacked along the channel dimension to get a total of four channels. Hence, the input to the DNN consists of a three-dimensional array with axes corresponding to segments (time), frequency bins, and channels. The frequency bandwidth of the input is chosen to be *twice* the bandwidth of the widest possible signal in an all-sky search as defined in Table I. Therefore, with a segment FFT resolution of 1/6 h and bandwidth of the widest signal ~18.7 mHz, the DNN has a total input bandwidth of 824 frequency bins or ~38 mHz.[1]

---

[1]In the directed search case for Cassiopeia A, the bandwidth of signals at frequencies 500 and 1000 Hz is more than the maximum bandwidth of a signal in our all-sky searches. Therefore, we have not performed a search at those frequencies. This is also the case for any search at 1500 Hz.

The DNN input bandwidth is chosen to be twice the bandwidth of the widest possible signal to simplify the application of our DNN to a search over a wider frequency band. Similar to [13,14], this is done by sliding the DNN input window over the band with a half overlap between consecutive windows. Such a placement of input windows ensures that any signal in the frequency band of the search lies completely inside at least one input window.

### B. Network architecture

We have experimented with several different variations of the architecture presented in our previous work on targeted searches [12] to make it more suitable to the case of wide parameter-space searches. The final architecture that we have used can be divided into three blocks: the *stem block*, the *intermediate block*, and the *output block*.

The *stem block* contains the convolutional layers that have large kernel sizes along the frequency axis motivated by the principle of having kernels wide enough to contain the entire signal within a segment as discussed in Sec. IV A of [12]. In order to adapt this principle to a wide parameter search, we have modified these layers by introducing branches, each with a different kernel size, as used in the Inception model in GoogLeNet [18]. The different kernel sizes enable the DNN to better learn signals with different bandwidths as encountered in a wide parameter-space search.

Similar to [12], we first perform a one-dimensional convolution along the frequency axis with the widest kernel of 56 frequency bins and then a two-dimensional convolution with the widest kernel of $2 \times 24$ bins as shown in Table IV, where both choices of kernel size are motivated by the bandwidth of widest possible signal within one or two segments. The two-dimensional convolutional layer has, in each of its branches, another convolutional layer with a kernel size of $1 \times 1$. This additional layer reduces the number of channels from 256 to 32. This *dimensional reduction* is applied in order to reduce the computational

TABLE IV. The architecture of the stem block and the intermediate block of our DNN. The output shape of each layer or a set of layers is expressed as (T × F × C), which corresponds to the axis along segments (time), frequency, and channels, respectively. The convolutional layers are expressed in terms of their parameters (*kernel-size*, *strides*, *No. of channels*). Layers in one residual block are represented by putting them together in one square bracket (as done in Table 1 of [19]).

| Block | Output shape (T × F × C) | Layer | | | |
|---|---|---|---|---|---|
| Stem block | ... | Input layer (40 × 824 × 4) | | | |
| | 40 × 206 × 256 | Branch 1 | Branch 2 | Branch 3 | Branch 4 |
| | | 1 × 14, 1 × 4, 64 | 1 × 28, 1 × 4, 64 | 1 × 42, 1 × 4, 64 | 1 × 56, 1 × 4, 64 |
| | | Concatenate layer | | | |
| | 20 × 52 × 256 | Branch 1 | Branch 2 | Branch 3 | Branch 4 |
| | | 1 × 1, 1 × 1, 32 | 1 × 1, 1 × 1, 32 | 1 × 1, 1 × 1, 32 | 1 × 1, 1 × 1, 32 |
| | | 2 × 6, 2 × 4, 64 | 2 × 12, 2 × 4, 64 | 2 × 18, 2 × 4, 64 | 2 × 24, 2 × 4, 64 |
| | | Concatenate layer | | | |
| Intermediate block | 20 × 52 × 512 | Residual blocks: $\begin{bmatrix} 1 \times 1, 1 \times 1, 128 \\ 5 \times 5, 1 \times 1, 128 \\ 1 \times 1, 1 \times 1, 512 \end{bmatrix} \times 2$ | | | |
| | 10 × 26 × 1024 | 2 × 2, 2 × 2, 1024 | | | |
| | 10 × 26 × 1024 | Residual blocks: $\begin{bmatrix} 1 \times 1, 1 \times 1, 256 \\ 5 \times 5, 1 \times 1, 256 \\ 1 \times 1, 1 \times 1, 1024 \end{bmatrix} \times 2$ | | | |
| | 5 × 13 × 2048 | 2 × 2, 2 × 2, 2048 | | | |
| | 5 × 13 × 2048 | Residual blocks: $\begin{bmatrix} 1 \times 1, 1 \times 1, 512 \\ 5 \times 5, 1 \times 1, 512 \\ 1 \times 1, 1 \times 1, 2048 \end{bmatrix} \times 2$ | | | |

cost of the convolutions without losing any spatial information (used in [18]).

The *intermediate block* contains blocks of 2D convolutional layers with residual connections. These *residual blocks* are constructed according to the *bottleneck* design introduced in [19]. Each residual block contains three convolutional layers with kernel sizes of $1 \times 1$, $5 \times 5$, and $1 \times 1$, respectively, each with a different number of channels as shown in Table IV. The first $1 \times 1$ convolution reduces and the final $1 \times 1$ convolution increases the number of channels of the input. The $5 \times 5$ convolution is performed with a reduced input and output dimension i.e., the *bottleneck*. A layer normalization as described in [20] is performed on the linear output of the middle layer before the nonlinear activation. We employ multiple such residual blocks which are interspersed by convolutional layers that reduce the spatial dimension of the input as shown in Table IV.

The *output block* contains three layers: a global average pooling layer, a fully connected layer with 64 units, and a final output layer consisting of a single unit. The application of a global average pooling layer is not strictly consistent with the design principles outlined in Sec. IV A of [12]. But we have empirically found that using that instead of a flattened layer (reshaping input to a one-dimensional array) is more computationally efficient and also leads to better learning.

Except for the global average pooling and the final output layer, we use a ReLU activation function on all the other layers of the DNN. On the output layer, we apply a sigmoid activation function and produce a probability $\hat{y} \in [0, 1]$ of the data containing a signal. We use the sigmoid output of the DNN as it is well-suited for training. But, it is susceptible to numerical overflow and underflow when it is used as a detection statistic. This was also observed in previous works [12,13,21]. Therefore, we use the linear output of the final layer as a detection statistic.

The number of trainable parameters of our DNN architecture is ∼78 M and it requires ∼300 MB of memory per sample. The training was performed in NVIDIA A100-SXM4 GPUs with 40 GB of memory. The DNN was implemented in TENSORFLOW 2.0 ([22]) with the Keras API ([23]). We used the Weights and Biases platform ([24]) to monitor training and log losses and metrics during training.

### C. Estimation of $p_{\text{fa}}$

As mentioned in Sec. II, we estimate the sensitivity of a search method by calculating the detection probability $p_{\text{det}}$ at a fixed value of false-alarm probability $p_{\text{fa}}$. The calculation of $p_{\text{det}}$ is done in two steps: estimation of detection threshold on the DNN statistic at fixed $p_{\text{fa}}$ and calculation of $p_{\text{det}}$ by counting the number of threshold crossings in a set of signal samples.

In order to make a fair comparison with the results of the WEAVE matched-filter search and the DNN search in [13], we need to calculate $p_{\text{det}}$ at the fixed $p_{\text{fa}} = 1\%$ per $\Delta f_{\text{WEAVE}} = 50 \text{ mHz}$ band. But, our DNN input bandwidth is $\Delta f_{\text{DNN}} \sim 38 \text{ mHz}$. Thus, we need to estimate the right value of $p_{\text{fa}}$ (and therefore the threshold) per $\Delta f_{\text{DNN}}$ that corresponds to a $p_{\text{fa}} = 1\%$ per 50 mHz.

We can do so by using the following equation that relates the $p_{\text{fa}}$ calculated over $N$ *independent* bands $[p_{\text{fa}}(N)]$ to the $p_{\text{fa}}$ calculated over a single band $[p_{\text{fa}}(1)]$:

$$p_{\text{fa}}(N) = 1 - [1 - p_{\text{fa}}(1)]^N. \quad (3)$$

We see that $\Delta f_{\text{WEAVE}} = N \times \Delta f_{\text{DNN}}$ with a noninteger $N$. Assuming that Eq. (3) is valid for a noninteger $N$, we can use it to *naively* estimate the $p_{\text{fa}}$ per $\Delta f_{\text{DNN}}$. This gives us a value of $p_{\text{fa}} = 0.76\%$ per $\Delta f_{\text{DNN}}$. But, the value of $p_{\text{det}}$ calculated using the above method might not be accurate for $p_{\text{fa}} = 1\%$ per 50 mHz.

To properly cover a wide frequency band, the DNN input windows must be placed with a half overlap as described in Sec. III A. In this case, the assumption that noise in different input windows is mutually independent is not true. So, we cannot use Eq. (3) to compute an accurate equivalent $p_{\text{fa}}$ for $\Delta f_{\text{DNN}}$.

In order to get an accurate estimation of $p_{\text{fa}}$, we compare $p_{\text{fa}}$ computed over a much wider frequency band $\Delta f_{\text{wide}} \sim 1$ Hz. The exact value of $\Delta f_{\text{wide}}$ is chosen to be an integer multiple of $\Delta f_{\text{DNN}}$. We calculate the $p_{\text{fa}}(\Delta f_{\text{wide}})$ for a WEAVE search, using Eq. (3) applied to $N = \Delta f_{\text{wide}}/\Delta f_{\text{WEAVE}}$, as the assumption of mutually independent noise *is* valid for nonoverlapping WEAVE frequency bands.

We then estimate the threshold on the DNN statistic over $\Delta f_{\text{wide}}$ at fixed $p_{\text{fa}}(\Delta f_{\text{wide}})$ using repeated sliding searches over pure Gaussian noise. The DNN statistic over $\Delta f_{\text{wide}}$ is the maximum over the DNN output calculated at each of the input windows covering the bandwidth $\Delta f_{\text{wide}}$. Each input window overlaps with two neighboring windows, except for the windows at each boundary which overlap with only one neighboring window. The use of a $\Delta f_{\text{wide}} \gg \Delta f_{\text{DNN}}$ helps mitigate any resulting boundary effects in the estimation of threshold. Using this threshold, we calculate that a $p_{\text{fa}} \sim 0.48\%$ per $\Delta f_{\text{DNN}}$ is equivalent to $p_{\text{fa}} = 1\%$ per 50 mHz, showing a significant difference from the naive estimate of $p_{\text{fa}} \sim 0.76\%$.

Thus, the more accurate estimation of $p_{\text{det}}$ at a fixed $p_{\text{fa}} = 1\%$ per 50 mHz band allows a fair comparison with matched-filter searches. This method is more computationally expensive; therefore, we only use it to estimate $p_{\text{det}}$ once the training is complete, while we use the naive estimate during training and validation.

### D. Training and validation

We train 14 DNNs with the architecture described above for different search cases: six all-sky search cases and eight directed search cases. Each DNN is trained on a set of *signal* samples, i.e., signals added to samples of Gaussian noise, and an equal number of *noise* samples, i.e., pure Gaussian noise samples, all of the Gaussian noise samples being dynamically generated. The *training* dataset of *signal* samples consists of a set of 32 768 precomputed signals with parameters chosen randomly in the range described in Sec. II according to the search case. The dynamically generated noise, i.e., generating a new independent noise sample for every training step, prevents the DNN from overfitting to the features of a particular noise realization.

Each signal is injected into a noise realization at a fixed sensitivity depth $\mathcal{D}_{\text{MF}}^{90\%}$ corresponding to the search case shown in Table III. In one of the all-sky search cases, the DNN is trained on signals with frequency sampled from the entire search band of 20–1000 Hz. Since there is no corresponding $\mathcal{D}_{\text{MF}}^{90\%}$ given in [13], signals are injected at $\mathcal{D}_{\text{MF}}^{90\%}$ corresponding to the 1000 Hz case.

We also shift the entire signal along the frequency axis by a different random number of bins every step of training before adding it to noise. The number of bins by which the signal is shifted is chosen such that the shifted signal does not leave the DNN bandwidth. The purpose of doing this is to train the DNN to identify a signal at any position in the bandwidth of ∼38 mHz. Introducing such a random frequency shift while training is a way of augmenting our training dataset by simulating a larger population of signals.

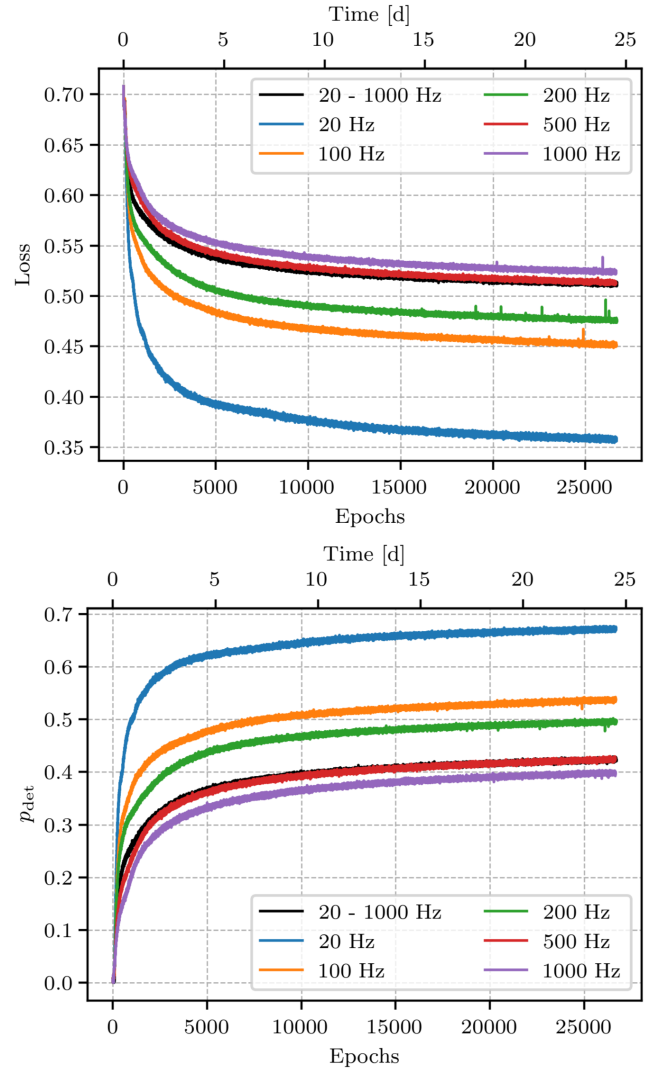

FIG. 1. Training loss (top) and $p_{\text{det}}$ (at fixed $p_{\text{fa}} = 0.76\%$ per $\sim 38$ mHz) (bottom) vs number of training epochs and time for the all-sky search cases.

We use the Adam optimizer [25] with a batch size of 256 samples for training. The loss function is a binary cross-entropy loss function, commonly used for classification tasks:

$$\mathcal{L}(y, \hat{y}) = \frac{1}{N} \sum_{i=1}^{N} [-y^i \log \hat{y}^i - (1 - y^i) \log(1 - \hat{y}^i)], \quad (4)$$

where $\hat{y}^i \in [0, 1]$ is the sigmoid output for the *ith* sample, $y^i$ is the corresponding label (0 for a noise sample and 1 for a signal sample), and $N$ is the total number of samples in a batch. Every epoch, we also compute detection probability $p_{\text{det}}$ at fixed $p_{\text{fa}} = 0.76\%$ per DNN bandwidth as a metric to evaluate the performance of the DNN. The loss and $p_{\text{det}}$ as a function of number of training epochs and time for the six all-sky search cases is shown in Fig. 1.

Every 50 epochs, we perform a validation step, in which we evaluate the loss and $p_{\text{det}}$ at $p_{\text{fa}} = 0.76\%$ per DNN input bandwidth on an independent dataset, called the *validation* dataset, consisting of a different sample of signals drawn from the same distribution as the *training* dataset. This evaluation on a dataset different from the *training* dataset provides an unbiased estimate to judge the performance of the DNN. We stop training once the training loss and $p_{\text{det}}$ start to plateau upon visual inspection as shown in Fig. 1. We have trained each of the DNNs for ∼25 days.

## IV. RESULTS AND DISCUSSION

In this section, we present the results of the tests that we have performed using trained DNNs on independent datasets. We show their capability to generalize over signal strength, frequency, and sky position. We also test their sensitivity as a function of sky position. We compare their performance with the previous best DNN sensitivity presented in [13].

### A. Performance on an independent test dataset

DNNs are susceptible to overfitting to the *training* and *validation* datasets. We therefore evaluate their performance on an independent dataset or *test* dataset. This *test* dataset contains an independent set of signals with their parameters drawn from the same distribution as that of the *training* and *validation* datasets. We compare the $p_{\text{det}}$ at fixed $p_{\text{fa}} = 1\%$ per 50 mHz (see Sec. III C) evaluated on signals injected at sensitivity depth of $\mathcal{D}_{\text{MF}}^{90\%}$ given in Table III. We find that $p_{\text{det}}$ evaluated on the *test* dataset matches that of the *training* and *validation* datasets (at the same $p_{\text{fa}}$) in each of the cases. Thus, the trained DNNs did not overfit to the set of signals in the *training* dataset, but generalized well to the unknown signals.

Tables V and VI show the comparison between sensitivity of our trained DNNs and the DNNs in the previous work [13] using two metrics: $p_{\text{det}}$ at fixed $p_{\text{fa}} = 1\%$ per 50 mHz band and $p_{\text{det}} = 90\%$ sensitivity depth ($\mathcal{D}^{90\%}$), respectively. For calculation of $p_{\text{det}}$, signals were injected into noise at $\mathcal{D}^{90\%}{}_{\text{MF}}$ given in Table III. We observe a general improvement in the sensitivity at all the frequencies, but the biggest improvement is seen at higher frequencies. The DNN trained on the entire search bandwidth, 20–1000 Hz also shows a very good performance, despite having to learn a much wider variety of signals structures.

TABLE V. Detection probabilities $p_{\text{det}}$ (at fixed $p_{\text{fa}} = 1\%$ per 50 mHz) along with the corresponding 95% confidence interval, evaluated on the *test* dataset for all-sky and directed searches compared to the values from the earlier work [13]. The $p_{\text{det}}$ is evaluated on a signal population injected into Gaussian noise at the corresponding $\mathcal{D}^{90\%}$ values listed in Table III. For the all-sky case of 20–1000 Hz, the reported value is the average of $p_{\text{det}}$ over different frequencies calculated in Sec. IV C.

| | All sky | | Directed (G347) | | Directed (CasA) | |
|---|---|---|---|---|---|---|
| | This work | Prev. work | This work | Prev. work | This work | Prev. work |
| $p_{\text{det}}(\%)$ | ($T_{\text{span}} = 10$ days) | ($T_{\text{span}} = 10^6$ s) | ($T_{\text{span}} = 10$ days) | ($T_{\text{span}} = 10^6$ s) | ($T_{\text{span}} = 10$ days) | ($T_{\text{span}} = 10^6$ s) |
| 20 Hz | $67.0^{+0.4}_{-0.4}$ | $60.5^{+3.7}_{-3.1}$ | $72.7^{+0.4}_{-0.4}$ | $71.2^{+3.1}_{-3.0}$ | $66.5^{+0.4}_{-0.4}$ | $54.6^{+3.3}_{-3.7}$ |
| 100 Hz | $52.8^{+0.4}_{-0.5}$ | $24.5^{+3.1}_{-3.1}$ | $68.2^{+0.4}_{-0.4}$ | ⋯ | $60.1^{+0.4}_{-0.4}$ | ⋯ |
| 200 Hz | $47.6^{+0.5}_{-0.4}$ | $11.2^{+3.1}_{-2.4}$ | $65.3^{+0.4}_{-0.4}$ | ⋯ | $57.3^{+0.5}_{-0.5}$ | ⋯ |
| 500 Hz | $41.5^{+0.4}_{-0.4}$ | $3.3^{+2.4}_{-1.3}$ | $62.0^{+0.4}_{-0.4}$ | $2.6^{+2.1}_{-1.2}$ | ⋯ | $0.6^{+0.6}_{-0.7}$ |
| 1000 Hz | $38.8^{+0.4}_{-0.4}$ | $0.7^{+0.7}_{-0.8}$ | $55.4^{+0.5}_{-0.5}$ | ⋯ | ⋯ | $0.7^{+1.0}_{-0.7}$ |
| 1500 Hz | ⋯ | ⋯ | ⋯ | $0.4^{+1.1}_{-0.6}$ | ⋯ | ⋯ |
| 20–1000 Hz | $36.0^{+0.4}_{-0.4}$ | ⋯ | ⋯ | ⋯ | ⋯ | ⋯ |

TABLE VI. $\mathcal{D}^{90\%}$ evaluated on the *test* dataset for all-sky and directed searches compared to the values from the earlier work [13] (not rescaled to time span of 10 days).

| $\mathcal{D}^{90\%}(/\sqrt{Hz})$ | All sky | | Directed (G347) | | Directed (CasA) | |
|---|---|---|---|---|---|---|
| | This work ($T_{span}$ = 10 days) | Prev. work ($T_{span}$ = $10^6$ s) | This work ($T_{span}$ = 10 days) | Prev. work ($T_{span}$ = $10^6$ s) | This work ($T_{span}$ = 10 days) | Prev. work ($T_{span}$ = $10^6$ s) |
| 20 Hz | 29.3 | 29.6 | 33.7 | 33.9 | 31.6 | 28.1 |
| 100 Hz | 23.9 | 17.5 | 31.8 | ⋯ | 29.3 | ⋯ |
| 200 Hz | 22.4 | 13.9 | 30.7 | ⋯ | 28.2 | ⋯ |
| 500 Hz | 20.1 | 9.7 | 28.7 | 11.7 | ⋯ | 0.0 |
| 1000 Hz | 18.3 | 7.9 | 26.5 | ⋯ | ⋯ | 1.4 |
| 1500 Hz | ⋯ | ⋯ | ⋯ | 1.3 | ⋯ | ⋯ |
| 20–1000 Hz | 18.8 | ⋯ | ⋯ | ⋯ | ⋯ | ⋯ |

### B. Generalization in signal strength

We studied the performance of our trained DNNs at different signal strengths by evaluating $p_{det}$ at fixed $p_{fa}$ = 1% per 50 mHz band on data containing signals injected at different values of $\mathcal{D}$. We show the plot of $p_{det}$ as a function of the injection sensitivity depth $\mathcal{D}$ for the case of all-sky search at 20 Hz in Fig. 2. The $p_{det}$ for the WEAVE matched-filter search was estimated by using the procedure described in [17] using the false-alarm threshold and mismatch in [13]. This dependence of $p_{det}$ on the $\mathcal{D}$ of signal injection is very similar to the curve for the WEAVE search in [13], but shifted to the left. This behavior shows that the DNN has not overfit to signals of sensitivity depth $\mathcal{D}^{90\%}$ used in training and generalizes to signals of unknown sensitivity depths as expected.

### C. Generalization in frequency

In this section we compare the performance of our DNNs trained on all-sky search cases at different frequencies by evaluating $p_{det}$ as a function of frequency of injected signals. For this purpose, we create all-sky search datasets at different frequencies across our search band 20–1000 Hz. They have the same signal parameters as given in Table I, except that the frequency of the signals is equal to the reference frequency of the dataset. Each of them contains 32768 precomputed signals similar to the *training* dataset. $\mathcal{D}^{90\%}$ of the injected signals at each frequency was linearly interpolated from the $\mathcal{D}^{90\%}$ values in Table III.

The 95% credible interval on $p_{det}$ evaluated on the above datasets as a function of frequency is shown in Fig. 3. We see the expected behavior that the DNN trained on the frequency specific datasets described in Table I perform their best at the frequency on which it was trained. Their performance gradually reduces as the frequency of signals move further away from their training frequency. This drop in performance occurs more steeply for DNNs trained on lower-frequency signals such as 20 Hz, while the one trained on higher frequency such as 1000 Hz shows a much more gradual drop in performance. This behavior indicates that the DNNs trained on signals with higher frequency generalize better to unknown frequencies than the ones trained on lower frequencies.

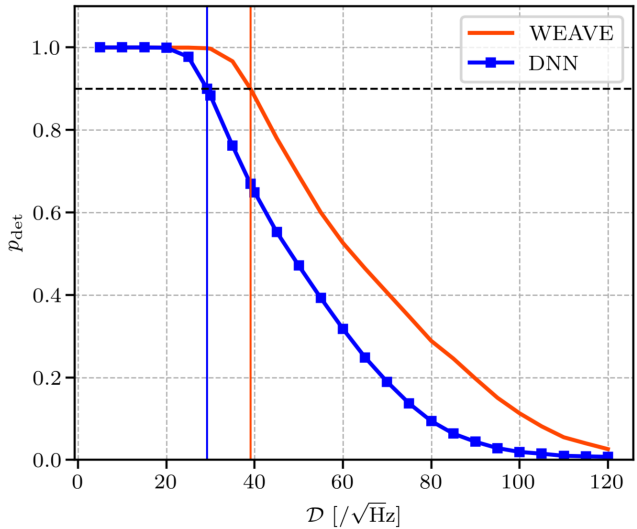

FIG. 2. Detection probability $p_{det}$ (at fixed $p_{fa}$ = 1% per 50 mHz band) versus signal injection depth $\mathcal{D}$ for trained DNN compared to a matched-filter search using WEAVE (in [13]) rescaled to time span of 10 days for the all-sky dataset at 20 Hz.

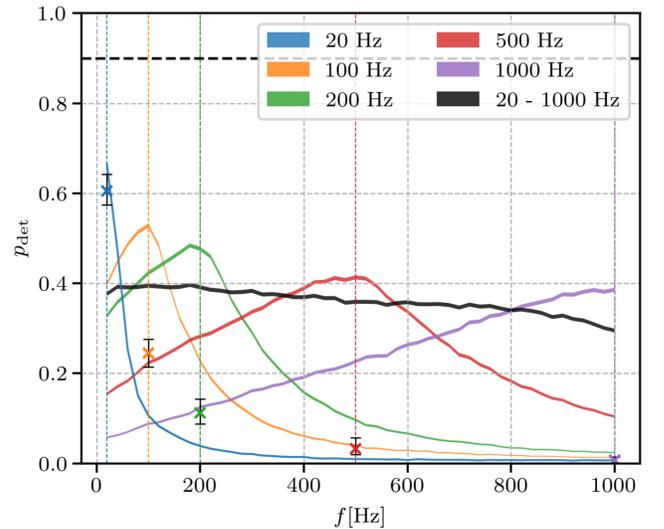

FIG. 3. 95% credible interval of the detection probability $p_{det}$ (at fixed $p_{fa}$ = 1% per 50 mHz band) as a function of the reference frequency for the DNNs trained on the six all-sky datasets. The dashed vertical lines mark the frequency in the same color of the five frequency-specific DNNs. The horizontal dashed black line represents performance of a coherent matched-filter search. The crosses with the black error bars denote the value of $p_{det}$ for the corresponding case from the earlier work [13].

Figure 3 also shows that the DNN trained on signals from the entire search band of 20–1000 Hz performs well on signals of all frequencies. It is the second-best-performing DNN at each of our chosen representative frequencies. Despite being only trained on 32768 signals, a small number over a very large band, it is able to generalize well to unknown signals. Moreover, we found that training this DNN using more signals, 65536 or 131072, does not lead to a significant improvement in its performance. Thus, a set of 32768 signals seems to be enough to learn the morphology of signals and generalize to a wide parameter space of an all-sky search.

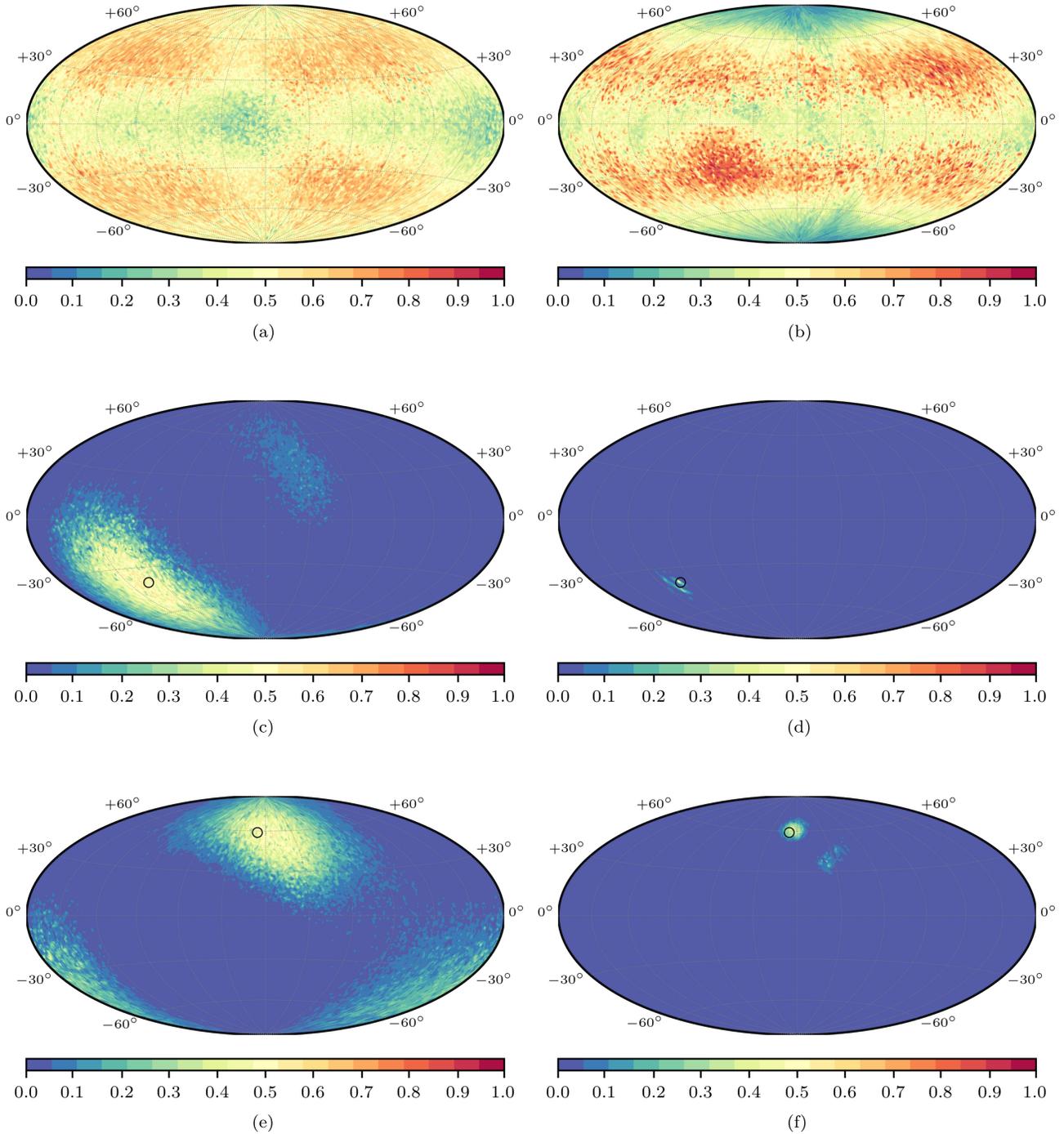

FIG. 4. Detection probability $p_{\text{det}}$ (at fixed $p_{\text{fa}} = 1\%$ per 50 mHz band) as a function of sky position of the injected signal in equatorial coordinates (Hammer projection). $p_{\text{det}}$ is measured at a fixed injection signal power $\rho^{50\%}$ yielding a $p_{\text{det}} = 50\%$ over the sky for the all-sky cases (a) All-sky at 20 Hz and (b) All-sky at 1000 Hz, and at the respective sky position for the directed cases (c) G347 at 20 Hz, (d) G347 at 1000 Hz, (e) CasA at 20 Hz, and (f) CasA at 200 Hz respectively.

Thus, an all-sky search across the entire search band of 20–1000 Hz can be performed using this single trained DNN. In order to do that, the DNN has to be evaluated at ∼50 000 input windows across the band, which takes ∼27 s using a batch size of 256. We see that the computational cost of performing the search is a negligible fraction of the cost of training the DNN.

The plot in Fig. 3 also shows the corresponding values of $p_{det}$ at fixed $p_{fa} = 1\%$ per 50 mHz for the previous best DNNs from [13], providing a concise visual illustration of the improvements in performance in our work.

### D. Generalization in sky position

In this section, we study the dependence of sensitivity of a DNN search on sky position of the signal.

We evaluated $p_{det}$ at 32 768 different sky positions with 200 signal injections each at a fixed $\rho^2$ [see Eq. (2)] such that the DNN had a given $p_{det} = 50\%$ over the entire sky for the all-sky cases and at the respective sky position for the directed cases. The rest of the signal parameters of these injections are the same as given in Table I. We present the results for two of the all-sky search cases and two from each of the directed search cases in the form of sky maps, as shown in Fig. 4. The values on a grid of sky points are obtained by interpolating between the sampled sky points.

In the all-sky search cases in Fig. 4, we see that the DNN is not uniformly sensitive over the entire sky. At 20 Hz, it is less sensitive in two spots near the celestial equator and more sensitive away from it, and at 1000 Hz, it is less sensitive at the celestial poles, but highly sensitive in the area between the poles and the equator. From the results in this paper as well as [12], we know that DNNs learn the structure of a narrow signal easier than a comparatively wider signal. This is a possible explanation for the variations in the sensitivity over sky position. Similarities in signal shapes over certain sky positions can also lead to an improvement in the sensitivity at those positions, as those signal shapes will be over-represented in the training set.

In the directed search cases in Fig. 4, we see that the sensitivity of the DNN at the sky position of the corresponding directed search is as expected. For the cases at low frequency, we see that it is sensitive in a wide region around the sky position of the corresponding directed search, but at high frequencies, it is sensitive in a much smaller region. The reason for this behavior is the frequency modulation due to the Doppler effect. At low frequencies, the shape of the signal changes slowly with sky position; therefore, we see such wide regions of the sky in which the DNN is sensitive. At higher frequencies, the shape of the signal changes rapidly with sky position; therefore, it is sensitive only in a very small region around the sky position of the corresponding directed search.

## V. CONCLUSIONS

In our previous work [12] we introduced new design principles for DNNs to perform CW searches. We also presented a novel architecture based on the design principles which was well suited for a targeted search, i.e., learning a single signal shape.

In this work, we present a new architecture based on appropriate updates to the design principles to make the architecture well suited for wide parameter-space searches of CWs. We compare the performance of the new architecture with the DNNs in [13]. We have devised a more accurate method of calculating $p_{det}$ at fixed $p_{fa}$ over a frequency band for a fair comparison with matched-filter searches. While the performance of the DNN search is not as good as that of the coherent matched-filter search, we have shown considerable improvement over previous DNN sensitivity [13]. We have also illustrated that a *single* DNN can be trained to perform a wide parameter search over the large search band of 20–1000 Hz with a sensitivity comparable to all the representative frequency cases.

Further improvements to the performance of a DNN search may be done by studying new architectures and training methodology such as the transformers [26,27]. Extending the time span of the search up to ∼1 yr is an important goal in order to scale the search to more realistic cases. Finally, training a DNN capable of estimating the parameters of detected signals can work toward developing a search method based entirely on deep learning.

## ACKNOWLEDGMENTS

This work has utilized the ATLAS computing cluster at the MPI for Gravitational Physics, Hannover, and the HPC system Raven at the Max Planck Computing and Data Facility.